# PCA-based lung motion model

R Li[*1], J Lewis[1], X Jia[1], T Zhao[2], J Lamb[2], D Yang[2], D Low[2], SB Jiang[1]

(1) University of California, San Diego, CA
(2) Washington University School of Medicine, St. Louis, MO

**Abstract**

*Organ motion induced by respiration may cause clinically significant targeting errors and greatly degrade the effectiveness of conformal radiotherapy. It is therefore crucial to be able to model respiratory motion accurately. A recently proposed lung motion model based on principal component analysis (PCA) has been shown to be promising on a few patients. However, there is still a need to understand the underlying reason why it works. In this paper, we present a much deeper and detailed analysis of the PCA-based lung motion model. We provide the theoretical justification of the effectiveness of PCA in modeling lung motion. We also prove that under certain conditions, the PCA motion model is equivalent to 5D motion model, which is based on physiology and anatomy of the lung. The modeling power of PCA model was tested on clinical data and the average 3D error was found to be below 1 mm.*

## Introduction

Respiration-induced organ motion is one of the major uncertainties in lung cancer radiotherapy, which may cause clinically significant targeting errors and greatly degrade the effectiveness of conformal radiotherapy. It is therefore crucial to be able to accurately model the respiratory motion. We distinguish between two main categories of respiratory motion models: one that devotes solely to the study of the motion of a single point (usually tumor center of mass); and the other one that attempts to model the motion of the entire lung by employing the spatial relations among different regions of the lung. This work will focus on the latter. Most works in the literature [1] rely more or less on the assumption of regular breathing and can yield suboptimal solutions if irregular breathing occurs.

As far as we know, there exist two spatiotemporal models which do not assume regular breathing patterns. Low *et al* [2] described a 5D lung motion model parameterized by tidal volume and airflow measured with spirometry, which allows characterization of hysteresis and irregular breathing patterns. More recently, Zhang *et al* [3] applied principal component analysis (PCA) to the 3D deformation field derived from a deformable image registration between a reference phase and other phases in a four-dimensional computed tomography (4DCT) data set. Although the PCA motion model in [3] seems promising for a small number of patients, there is still a need to understand the underlying reason why it works and whether there is any connection between the two lung motion models. In this paper, we will present a much deeper and detailed analysis of the PCA-based lung motion model. We provide the theoretical justification of the effectiveness of PCA in modeling lung motion. We shall see that it is closely related to Low's physiological 5D lung motion model and that under certain conditions, these two models are actually equivalent.

## Material and methods

**Construction of the PCA lung motion model**

We first briefly describe how PCA may be used to construct a lung motion model. We form a matrix $\mathbf{X}$, where each row represents the displacement vectors of a certain voxel in the lung along one of the three coordinates in space at all time points. If we perform PCA on the covariance matrix of $\mathbf{X}$, we will get a set of eigenvectors $\mathbf{u}_1, \mathbf{u}_2, ...$, corresponding to a set of non-negative eigenvalues $\lambda_1, \lambda_2, ...$ . Intuitively, each eigenvalue represents how much variation or variance in the data is captured by the corresponding eigenvector. In practice, the eigenvalue usually decreases very fast. The hypothesis is that every possible lung motion state $\mathbf{x}(t)$ can be approximated by a linear combination of the eigenvectors corresponding to the largest eigenvalues, i.e.,

$$\mathbf{x}(t) \approx \overline{\mathbf{x}} + \sum_{k=1}^{K} \mathbf{u}_k w_k(t)$$

where $\overline{\mathbf{x}}$ is the sample mean of all the columns in matrix $\mathbf{X}$, and scalars $w_k(t)$ are called PCA coefficients.

**Properties of PCA motion model**

An important property of PCA is that it provides the best linear representation of the data in the least mean-square (LMS) sense. Specifically, of all the $K$-dimensional linear approximation of the deformation vector fields, the mean square error is minimized by summing the time-averaged deformation vector fields and a linear combination of the first $K$ eigenvectors (with the $K$ largest eigenvalues) weighted by the corresponding principal components.

Another property concerns with the implicit regularization imposed by the PCA motion model. Let $\tilde{\mathbf{Y}} = \mathbf{U}\mathbf{\Lambda}\mathbf{W}^T$ be a low-rank approximation to the deformation vector fields, where $\mathbf{\Lambda}$ is a $K$ by $K$ diagonal matrix containing the first $K$ largest eigenvalues ($K \leq rank(\tilde{\mathbf{X}})$), $\mathbf{U}, \mathbf{W}$ are unitary matrices with size $N$ by $K$ and $M$ by $K$, where each column is the corresponding left and right eigenvector, respectively. If we look at one particular row $\tilde{\mathbf{y}}_i = \mathbf{u}_i \mathbf{\Lambda} \mathbf{W}^T$, it is the motion of the corresponding voxel along one direction over time. We can rewrite as: $\mathbf{u}_i = \tilde{\mathbf{y}}_i \mathbf{W} \mathbf{\Lambda}^{-1}$. If we look at the difference between any two rows (i.e., motion of two voxels along one direction), we can see that,

$$\|\Delta \mathbf{u}\|^2 = \sum_{i=1}^{K} \frac{(\Delta \tilde{\mathbf{y}} \cdot \mathbf{w}_i)^2}{\lambda_i^2} \leq \left(\sum_{i=1}^{K} 1/\lambda_i^2\right) \|\Delta \tilde{\mathbf{y}}\|^2,$$

where $\mathbf{w}_i$ are the column vectors of $\mathbf{W}$.

But $\|\Delta \tilde{\mathbf{y}}\|^2 \leq \|\Delta \tilde{\mathbf{x}}\|^2$, we have: $\|\Delta \mathbf{u}\|^2 \leq \left(\sum_{i=1}^{K} 1/\lambda_i^2\right) \|\Delta \tilde{\mathbf{x}}\|^2$.

The implication is that if two voxels move similarly, then their motion represented by PCA will also be similar, provided that the principal components kept in the model do not have vanishing eigenvalues associated with them. Without this property, the magnitude of the difference between two eigenvectors is not limited by the corresponding difference between the real motion and can be arbitrarily large, then the motion of two voxels reconstructed by PCA can be wildly different, even if they move very similarly, which is not desirable.

**Respiratory phantom with $\cos(t)$ motion**

The main feature of respiration is that it is somewhat (though not perfectly) periodic. The simplest function that captures this feature is a cosine function. In the cosine respiratory phantom, the motion of each voxel along each of three coordinates in space is in the form of cosine functions. We allow arbitrary amplitude and arbitrary phase for each cosine function. In matrix form:

$$\mathbf{X} = \begin{bmatrix} A_1 \cos(\theta + \varphi_1) & A_1 \cos(2\theta + \varphi_1) & ... & A_1 \cos(M\theta + \varphi_1) \\ ... & ... & ... & ... \\ A_N \cos(\theta + \varphi_N) & A_N \cos(2\theta + \varphi_N) & ... & A_N \cos(M\theta + \varphi_N) \end{bmatrix}$$

where $\mathbf{X}$ is an $N$ by $M$ matrix; $N$ is the number of voxels in the lung times 3, and $M$ is the number of samples in time. $A_1, ..., A_N$ and $\varphi_1, ..., \varphi_N$ are amplitude and phase; $\theta$ is the time interval between successive samples.

It may seem to be an idealistic respiratory phantom at first. However, notice that any 3 rows in the above matrix is exactly the parametric form of an ellipse in 3D, so the 3D trajectory of each voxel follows an ellipse, and since we allow arbitrary amplitude and phase for each spatial coordinate, this ellipse (i.e., the 3D trajectory of a voxel) can have arbitrary shape, size and orientation in space. Therefore, this respiratory phantom can be seen as a coarse approximation to regular breathing.

**Respiratory phantom with $\cos^{2n}(t)$ motion**

In this respiratory phantom, we replace the cosine functions in the above phantom with even power of cosine functions, i.e., $X(t) = A\cos^{2n}(t + \Phi)$. We still allow arbitrary amplitude and arbitrary phase for each function. This formula has been used by Lujan et al [4] to model lung motion. The bias term in this function does not matter for PCA because of centering and is set to zero without loss of generality.

**Clinical data**

In this study, 2 patients were enrolled under an IRB-approved protocol and scanned using a 64-slice CT scanner (Philips 64-slice Brilliance CT) operating in ciné mode with a slice-thickness of 0.625 mm. Each contiguous set of the simultaneously acquired 64 CT slices was called a couch position, which covers 4 cm in the longitudinal axis of the patient. The scanner was operated to acquire 25 scans per couch position using a 0.42 sec rotation, 360° reconstruction, and 0.32 sec between successive ciné acquisitions, requiring 18.2 sec to acquire the 25 scans.

In order to get the displacement vectors for each voxel in the lung, we performed deformable image registration between a reference CT scan and all other scans at a particular couch position. The particular couch position we looked at is the second most inferior couch position, where the overall 3D motion is typically the largest in the thorax while a reliable registration may still be obtained.

## Results and discussion

**Results on respiratory phantom with $\cos(t)$ motion**

The main result is that using 2 PCA coefficients and eigenvectors will completely represent the lung motion under cosine phantom. First we show that the motion matrix **X** has a rank of 2. We can express **X** as a summation of 2 rank-1 matrices, each being an outer product of a column vector and a row vector.

$$\mathbf{X} = \begin{bmatrix} A_1 \cos\varphi_1 \\ \ldots \\ A_N \cos\varphi_N \end{bmatrix} [\cos\theta \quad \ldots \quad \cos M\theta]$$

$$- \begin{bmatrix} A_1 \sin\varphi_1 \\ \ldots \\ A_N \sin\varphi_N \end{bmatrix} [\sin\theta \quad \ldots \quad \sin M\theta]$$

Since $rank(\mathbf{A}+\mathbf{B}) \leq rank(\mathbf{A}) + rank(\mathbf{B})$ for 2 matrices of the same size, we know that $rank(\mathbf{X}) \leq 2$. In fact, $rank(\mathbf{X}) = 2$ with probability 1.

Appendix A derives the necessary and sufficient condition that PCA and 5D motion models are *equivalent* for the cosine respiratory phantom: $E[A^2 \sin(2\Phi)] = 0$, where $\Phi$ is a random variable representing the phase shift between tidal volume and voxel motion along each coordinate. If $A, \Phi$ are uncorrelated random variables, the equivalence condition becomes $E[\sin(2\Phi)] = 0$. There are many situations where this condition is satisfied. For instance, any distribution of $\Phi$ that is symmetric about 0, or $\pm\pi/2$, or $\pm\pi$ will satisfy the condition and yields equivalent PCA and 5D lung motion models.

### Results on respiratory phantom with $\cos^{2n}(t)$ motion

We first show that in the discrete case, $rank(\mathbf{X}) = 2n+1$. First notice that $\cos^{2n} t = \sum_{k=0}^{n} c_k \cos(2kt)$, where $c_k$ are known constants. Then the motion matrix is can be written as:

$$\mathbf{X} = \sum_{k=0}^{n} c_k \cdot \begin{bmatrix} A_1 \cos 2k\varphi_1 \\ \ldots \\ A_N \cos 2k\varphi_N \end{bmatrix} [\cos 2k\theta \quad \ldots \quad \cos 2kM\theta]$$

$$- \sum_{k=1}^{n} c_k \cdot \begin{bmatrix} A_1 \sin 2k\varphi_1 \\ \ldots \\ A_N \sin 2k\varphi_N \end{bmatrix} [\sin 2k\theta \quad \ldots \quad \sin 2kM\theta]$$

We can see that matrix **X** is a summation of $2n+1$ rank-1 matrices. So $rank(\mathbf{X}) = 2n+1$. One can show that the rank of $\tilde{\mathbf{X}}$ (centered matrix) is actually reduced by 1, *i.e.*, $2n$. So a full representation of **X** only requires at most $2n$ PCA coefficients and eigenvectors.

It is straightforward though tedious to show that the eigenfunctions are a linear combination of sinusoidal functions at even multiples (up to $2n$) of the fundamental frequency. In general, the actual form of the eigenfunctions depends on the distribution of the amplitude and phase for all the voxels. This is in contrast to the 5D model, where the "eigenfunctions" are external surrogates and do not depend on internal lung motion.

### Applications to clinical data

We validate the PCA lung motion model by its modeling power. We use the deformation vectors from all 25 scans as training data for PCA. Figure 1 shows the 3D positions of a particular voxel for all 25 scans in patient 1 as well as the interpolated 3D trajectory from PCA model output using 2 coefficients and eigenvectors. We can see that PCA model is able to capture the hysteresis component of the motion. Table 1 lists the average 3D root-mean-square (RMS) error for all voxels for both PCA and 5D lung motion model for 2 patients. For the PCA model, the modeling error decreases with the number of coefficients. But here we used 2 coefficients and eigenvectors to calculate the error for comparison with 5D model. The smaller error of PCA model is expected because of the way it is constructed.

There is a very high correlation between the first PCA coefficient (average: 0.97) and tidal volume as well as the first eigenvector and **α** (average: 1.00). On the other hand, the correlation between the second PCA coefficient and airflow as well as the second eigenvector and **β** is more variable (average of 0.48 and 0.66, respectively), which explains their different modeling power.

## Conclusion

We have presented a detailed analysis and validation of a lung motion model based on PCA. Given two types of respiratory phantoms based on simple and yet realistic assumptions, we established the theoretical result that a complete representation of the lung motion may require different numbers of PCA coefficients and eigenvectors depending the overall structures of lung motion. We applied the PCA motion model to clinical data and tested its modeling power. There are several clinical applications of the PCA motion model. Once lung motion is parameterized by a few PCA coefficients and eigenvectors, single-marker measurements can be used to track the motion of entire lung. Other tracking or imaging techniques may be used too. For instance, real-time surface imaging can then be used to derive the entire body (including lung) motion. By deforming a reference CT image with the deformation fields parameterized by PCA coefficients and match its projection with fluoroscopy, one can also obtain the entire lung motion in real time.

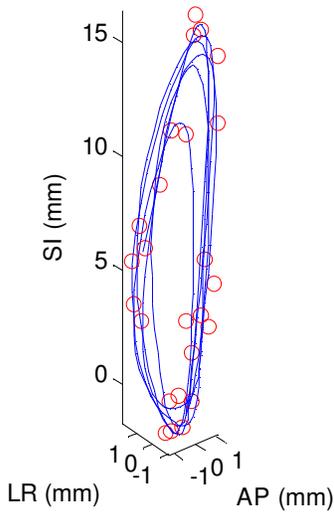

**Figure 1:** The 3D positions of a particular voxel for all 25 scans in patient 1 (shown in circles) as well as the interpolated 3D trajectory from PCA model using 2 PCA coefficients and eigenvectors.

| Patient Index | 5D model error (mm) | PCA model error (mm) | Mean motion magnitude (mm) |
|---|---|---|---|
| 1 | 0.71 | 0.46 | 11.4 |
| 2 | 0.64 | 0.39 | 9.0 |

**Table 1:** Average 3D RMS modeling error

**Appendix A**

In the following, we derive the necessary and sufficient conditions for 5D and PCA models to be equivalent. If we consider the values $A_i, \varphi_i$ in each row as realizations of random variables $A, \Phi$, as $\theta \to 0$ and $M\theta \to T$, the motion matrix **X** can be represented by a random process: $X(t) = A\cos(t+\Phi)$, where $0 \le t < T$.

The 5D lung motion model has the following expression: $\mathbf{x}(t) = \mathbf{x}_0 + \boldsymbol{\alpha} \cdot v(t) + \boldsymbol{\beta} \cdot f(t)$, where $v(t)$ is the tidal volume and $f(t)$ is the airflow. Suppose tidal volume is $v(t) = A_0 \cos t$. Then air flow is $f(t) = -A_0 \sin t$. Then 5D motion model is:

$$X(t) = \frac{A\cos\Phi}{A_0} \cdot v(t) + \frac{A\sin\Phi}{A_0} \cdot f(t). \quad (A-1)$$

PCA in the continuous domain becomes Karhunen-Loeve (KL) expansion. Karhunen-Loeve theorem [5] states that a random process can be represented by a linear combination of an infinite number of orthonormal deterministic functions with uncorrelated random coefficients, i.e., $X(t) = \sum_{k=1}^{\infty} Z_k \phi_k(t)$. The eigenfunctions $\phi_k(t)$ must satisfy: $\int_0^T K(s,t)\phi(s)ds = \lambda \cdot \phi(t)$.

$$K(s,t) = E[X(s)X(t)]$$
$$= \tfrac{1}{2} E\left[A^2 \left(\cos(s-t) + \cos(s+t+2\Phi)\right)\right]$$
$$= \gamma_1 \cdot \cos s \cos t + \beta \cos s \sin t +$$
$$\beta \sin s \cos t + \gamma_2 \cdot \sin s \sin t$$

where, $\beta = -\tfrac{1}{2} E[A^2 \sin(2\Phi)]$, $\gamma_1 = E[A^2 \cos^2 \Phi]$, $\gamma_2 = E[A^2 \sin^2 \Phi]$.

We express the eigenfunction in [0, *T*] in its Fourier expansion: $\phi(t) = \sum_{k=0}^{\infty} \left(a_k \cos(kt) + b_k \sin(kt)\right)$.

The eigenvalues and Fourier coefficients must satisfy:
$$\begin{cases} \lambda a_1 = \pi(\gamma_1 a_1 + \beta b_1) \\ \lambda b_1 = \pi(\beta a_1 + \gamma_2 b_1) \end{cases} \text{ and } \pi\left(a_1^2 + b_1^2\right) = 1.$$

Since $a_1, b_1$ can not be 0 at the same time, we have $\det C(\lambda) = 0$, where, $C(\lambda) = \begin{pmatrix} \gamma_1 - \lambda/\pi & \beta \\ \beta & \gamma_2 - \lambda/\pi \end{pmatrix}$.

*Sufficiency.*
If $\beta = 0$, then $\lambda_1 = \pi\gamma_1, \lambda_2 = \pi\gamma_2$. The corresponding eigenfunctions are: $\phi_1(t) = \tfrac{1}{\sqrt{\pi}} \cos t, \phi_2(t) = -\tfrac{1}{\sqrt{\pi}} \sin t$.

KL expansion gives: $X(t) = Z_1 \cdot \tfrac{1}{\sqrt{\pi}} \cos t + Z_2 \cdot \tfrac{1}{\sqrt{\pi}} \sin t$, where $Z_1 = \sqrt{\pi} A \cos\Phi, Z_2 = \sqrt{\pi} A \sin\Phi$.

We can see that except for a scaling factor, tidal volume and air flow are exactly the first and second eigenfunctions, and the last 2 parameters in 5D motion model are the corresponding PCA coefficients.

*Necessity.*
If $X(t)$ can be represented by (A-1), then KL theorem states that $E[Z_1 Z_2] = 0$, i.e., $E[A^2 \sin(2\Phi)] = 0$.
Q.E.D.

## References

[1] Zeng R, Fessler J A and Balter J M 2007 Estimating 3-D respiratory motion from orbiting views by tomographic image registration *IEEE Trans Med Imaging* **26** 153-63

[2] Low D A, Parikh P J, Lu W, Dempsey J F, Wahab S H, Hubenschmidt J P, Nystrom M M, Handoko M and Bradley J D 2005 Novel breathing motion model for radiotherapy *Int J Radiat Oncol Biol Phys* **63** 921-9

[3] Zhang Q, Pevsner A, Hertanto A, Hu Y C, Rosenzweig K E, Ling C C and Mageras G S 2007 A patient-specific respiratory model of anatomical motion for radiation treatment planning *Med Phys* **34** 4772-81

[4] Lujan A E, Larsen E W, Balter J M and Ten Haken R K 1999 A method for incorporating organ motion due to breathing into 3D dose calculations *Med Phys* **26** 715-20.

[5] Loeve M 1978 *Probability theory* vol 2 (New York: Springer-Verlag)